\documentclass[12pt]{article}
\usepackage{axodraw}

\newcommand{\beq}{\begin{equation}}
\newcommand{\eeq}{\end{equation}}
\newcommand{\beqa}{\begin{eqnarray}}
\newcommand{\eeqa}{\end{eqnarray}}
\newcommand{\beqar}{\begin{eqnarray*}}
\newcommand{\eeqar}{\end{eqnarray*}}

\newcommand{\al}{\alpha}
\newcommand{\be}{\beta}

\def\spa          {\ \ \ }
\def\non          {\nonumber}
\def\ha           {\mbox{$\frac{1}{2}$}}

\def\spa          {\ \ \ }
\def\mand         {\spa\mbox{and}\spa}

\def\Tr           {\mbox{\rm Tr}\,}
\def\STr          {\mbox{\rm STr}\,}
\def\Str          {\mbox{\rm Str}\,}

\def\cd           {{\cdot}}
\def\ran          {\rangle}
\def\lan          {\langle}
\def\fsH    {H\!\!\!\!/\,}
\newcommand{\del}{\delta}

\newcommand{\eps}{\epsilon}
\newcommand{\ga}{\gamma}

\newcommand{\inn}{\!\cdot\!}

\newcommand{\lam}{\lambda}

\newcommand{\z}{\zeta}

\newcommand{\ie}{{\it i.e.,}\ }
\newcommand{\labell}[1]{\label{#1}} 
\newcommand{\reef}[1]{(\ref{#1})}
\newcommand\prt{\partial}

\newcommand\ls{\ell_s}

\newcommand\cL{{\cal L}}
\newcommand\cD{{\cal D}}

\newcommand\bz{\bar{z}}

\newcommand\M[2]{M^{#1}{}_{#2}}

\begin{document}
\baselineskip 18pt%
\begin{titlepage}
\vspace*{1mm}%
\hfill
\vbox{

    \halign{#\hfil         \cr
         CERN-PH-TH/2010-034\cr
         IPM/P-2010/003  \cr
           } 
      }  
\vspace*{10mm}
\vspace*{12mm}%

\center{ {\bf \Large On effective actions of BPS branes
 and their higher derivative corrections
 
}}\vspace*{3mm} \centerline{{\Large {\bf  }}}
\vspace*{5mm}
\begin{center}
{Ehsan Hatefi}$\footnote{On leave of Ferdowsi University of Mashhad, School of Physics, 
Institute for research in \\fundamental sciences (IPM), P.O.Box 19395-5531, Tehran, Iran.\\
E-mail:ehsan.hatefi@cern.ch}$

\vspace*{0.8cm}{ {
Theory Group, Physics Department, CERN, CH-1211, Geneva 23, Switzerland}}
\vspace*{1.5cm}
\end{center}
\begin{center}{\bf Abstract}\end{center}
\begin{quote}
We calculate in detail the disk level S-matrix element of one Ramond-Ramond field and three gauge field vertex operators in the world volume of BPS branes, to find four gauge field couplings to all orders of $\alpha'$ up to on-shell ambiguity. Then using these infinite couplings we find that the massless pole of the field theory amplitude is exactly equal to the  massless pole S-matrix element of this amplitude for the $p=n$ case  to all orders of $\alpha'$. Finally we show that the infinite massless poles and the contact terms of this amplitude for the $p=n+2$ case can be reproduced by the Born-Infeld action and the Wess-Zumino actions and by their higher derivative corrections.
\end{quote}
\end{titlepage}
\section{Introduction}
A D$_p$-brane is a $(p+1)$-dimensional hypersurface in a space-time
defined such that open strings can end on it. They are
sources of Ramond-Ramond (P+1)-form fields in IIA and IIB string theories \cite{Polchinski:1995mt}. They have
been studied and many properties of
them have been investigated  \cite{Witten:1995im,Polchinski:1996na}, being the center of attention both in theory and phenomenology.
The stable D$_p$-branes ($p$ is even in IIA
 and odd in IIB theory) preserve half of
supersymmetry. This implies that the spectrum of open strings has 16 supersymmetries and has to be 
tachyon-free. An important point is that the
 supersymmetry transformation requires
 the momenta of the scattering states to be only along the brane
 directions. 
 
In fact,  stability, supersymmetry, conserved Ramond-Ramond (RR) charge and having no tachyons are all properties of these type II
D$_p$-branes. All supersymmetric D$_p$-branes in
IIA can be generated as bound states of D$_9$-branes \cite{hor1}.
They can also be derived
from K-theory~\cite{wit1}.

The world-volume theory of a D$_p$-brane involves a massless U(1) vector
$A_a$, $9-p$ real massless scalars $\phi^i$
which describe transverse oscillations of the brane
and their super partner fermions \cite{Dbrane}. At leading order, the low-energy action for these fields 
corresponds to the dimensional reduction of a ten-dimensional U(1) super-Yang Mills theory. There are higher order $\alpha'=l_s^2$
corrections, where $l_s$ is the string length scale. When derivatives
of the field strengths (and second derivatives of the scalars) are small on the string scale, then the
action to all orders in the field strength takes the Born-Infeld form \cite{Abouelsaood:1986gd,Leigh:1989jq} (also see
\cite{Tseytlin:1999dj}).

On the other hand, when there are N coincident D$_p$-branes, the U(1) gauge symmetry of a single
D$_p$-brane is enhanced to the non-abelian U(N) symmetry \cite{Witten:1995im}.
The action for constructing non-abelian D$_p$-branes on a general background was given by Myers  \cite{Myers:1999ps} and other authors \cite{Taylor:1999pr}. 

The low energy action describing the dynamics of  D$_p$-branes consists of two
parts.
The first part is Born-Infeld action 
\beq
{S}_{BI}=-T_p \int d^{p+1}\sigma\,\STr\left(e^{-\phi}\sqrt{-\det\left(
P\left[E_{ab}+E_{ai}(Q^{-1}-\delta)^{ij}E_{jb}\right]+
\lambda\,F_{ab}\right)\,\det(Q^i{}_j)}
\right),
\labell{finalbi}
\eeq
with 
\beq
E_{ab}=G_{ab}+B_{ab}
\qquad{\rm ,}\qquad
Q^i{}_j\equiv\delta^i{}_j+i\lambda\,[\Phi^i,\Phi^k]\,E_{kj},
\labell{extra6}
\eeq
where $\lambda=2\pi \ls^2$, $T_p$ is the brane tension, P[...] indicates
pull-back of background metric and NSNS two-form (a,b = 0, .., 9), $F_{ab}$ is
the field strength of gauge field and STr(...) is symmetric trace prescription. For more details see \cite{Myers:1999ps}. In addition BI action provides the kinetic terms for the world-volume fields, it
also contains the couplings of the D$_p$-brane to the massless Neveu-Schwarz fields in the
bulk. 
 The second part is the Wess-Zumino action, which contains the coupling of the U($N$) 
massless world volume vectors to the closed string RR
field (indicated by $C$) \cite{Polchinski:1995mt,li1996a}
\beqa
S_{WZ}&=&\mu_p \int_{\Sigma_{(p+1)}} C \wedge \Tr e^{i2\pi\alpha'F} \ .
\label{WZ'}
\eeqa
where $\Sigma_{(p+1)}$ is the world volume, $\Tr$ is over the
Chan-Paton factors, $\mu_p$ is the RR charge of branes and $F$ is the field strength of the gauge field which
defined $F=\frac{1}{2}F_{ab}dx^a\wedge dx^b$.
Using the Taylor expansion one finds various couplings including two gauge fields, three gauge fields and so on.
Therefore the effective theory has two parts 
\beqa 
S_{BPS}&=&S_{BI}+S_{WZ}\labell{bps}\nonumber\eeqa
One method for finding these effective actions is the BSFT. By applying this formalism we can find WZ couplings.
In this framework, it has been argued in \cite{Kraus:2000nj} when the RR field is constant, there is no higher derivative correction to the WZ couplings. The WZ term in this formalism is given by \cite{Kraus:2000nj,Takayanagi:2000rz}
\beqa
S_{WZ}&=&\mu_p' \int_{\Sigma_{(p+1)}} C \wedge \Str e^{i2\pi\alpha'\cal F},\labell{WZo}\eeqa 
 in which $\cal F$ is the curvature of superconnection.
To study WZ couplings for BPS branes we use the second approach, which is the S-matrix method. As we are working with Born-Infeld action in a flat background, we set $G_{ab}=\eta_{ab}$ and $B_{ab}=0$. The Born-Infeld action can then be written as  
\beqa
S_{BI}=-T_p \int d^{p+1}\sigma\,\STr\bigg(e^{-\phi}
\sqrt{-\det(\eta_{ab}
+2\pi\alpha'F_{ab}) }
\bigg). \label{BI}\eeqa

One important tool in string theory is scattering theory. In fact, the string theory
corrections to field theory may be found perturbatively in $\alpha'$ by means of scattering
amplitude arguments. In string theory, one can narrate the scattering of closed strings from a D$_p$-brane as follows. 

The string background is taken to be flat space, however, interactions of
 closed strings with a D$_p$-brane are described by world-sheets with boundary. The boundary
of world sheets must be fixed to the surface at the position of the D$_p$-brane. In fact, we must
consider Dirichlet boundary conditions on the fields transverse to the D$_p$-brane and Neumann boundary conditions on 
the fields along world volume of the D$_p$-brane \cite{Polchinski:1994fq}. The appearence of the D$_p$-brane physics has yielded
  a deep change in the significance of open strings, with some efforts for explaining them being related to
 \cite{Park:2007mc
 }. It 
 was conjectured that quantum effects of open strings moving on
D$_p$-branes will produce D$_p$-branes geometry. Some previous works on scattering that involved a D$_p$-brane
and some works about applications on D$_p$-branes can be found in \cite{Hashimoto:1996bf}. Firstly
 the massless vertex operators were constructed for
 the external open strings on the D$_p$-branes.
For the scattering of massless states the D$_p$-brane geometry
 is the extremal case. One can calculate the scattering of massless states from supersymmetric D$_p$-branes in type II
theory. 

T-duality transformation transforms D$_p$-branes from 
type IIA to type IIB and vice versa. Using T-duality transformation one can substitute a
scalar field with a gauge field and vice versa. Open string states on the D$_p$-brane do not have transverse components of their momenta. 
In this paper, we would like to use T-duality to find 
higher derivative couplings of four gauge fields from higher derivative couplings of four transverse scalar fields  \cite{{Garousi:2008xp}}.

The organization of the paper is as follows. In section 2 we calculate a tree-level four point string scattering including, one RR and three gauge field vertex operators in the world volume type II superstring theory and make a few remarks by studying direct computations. The world-volume theory can be described by  Berkovits's
 superstring field theory \cite{Berkovits:1995ab}. Also the world-volume theory may be
 rewritten in terms of massless fields and an infinite number of their derivatives.
In section 3 we examine the low energy limit by sending all Mandelstam variables to zero. In section 4 we consider
 low energy field theory to find desired couplings and then using these couplings we produce first massless pole 
 for $p=n$ case.
 
 To obtain the infinite massless
  poles for this case, one needs to know the higher derivative couplings of four gauge fields. To this aim, using T-duality transformation in section 5 we find higher derivative couplings of four gauge fields up to on-shell ambiguity. Then using these couplings in field theory we will show that the  massless poles of this S-matrix element are exactly reproduced to all orders of $\alpha'$. Finally in section 5.2 we obtain all infinite massless poles in field theory for $p=n+2$ case
and find a consistent result between string theory and field theory amplitudes. In addition
we generate contact terms of this amplitude in leading order and next to the leading order.

Before continuing our calculations, let us explain the conventions. Our index conventions are that lowercase
Greek indices take values in the whole ten-dimensional spacetime, e.g., $\mu,\nu$ = 0, 1,..., 9; early
Latin indices take values in the world-volume, e.g., $a, b, c$ = 0, 1,..., p; and middle Latin indices
take values in the transverse space, e.g., $i,j$ = p + 1,...,9. Thus, for example, $G_{\mu\nu}$ indicates
the entire spacetime metric, while $G_{ab}$ and $G_{ij}$ indicate metric components for directions parallel
and orthogonal to the D$_p$-branes, respectively.

\section{The four point superstring scattering (CAAA) }
In this section, using the conformal field theory technique we evaluate the string scattering amplitude to find all couplings of one closed string RR field to three gauge fields on the world-volume of a single BPS D$_p$-brane with flat empty space background. To calculate a S-matrix element, one firstly needs to choose the picture of the vertex operators. The sum of the superghost charges must be -2 for disk level amplitudes. 

A great deal of effort for the scattering amplitudes at tree level has been made
\cite{Kennedy:1999nn,Medina:2002nk,Garousi:2007fk,hatefi:2008,Stieberger:2009hq}.
The S-matrix element of one closed string RR field and three gauge fields 
is given by the following correlation function 
\begin{eqnarray}
{\cal A}^{CAAA} & \sim & \int dx_{1}dx_{2}dx_{3}dzd\bar{z}\,
  \lan V_{A}^{(0)}{(x_{1})}
V_{A}^{(0)}{(x_{2})}V_A^{(-1)}{(x_{3})}
V_{RR}^{(-\frac{1}{2},-\frac{1}{2})}(z,\bar{z})\ran,\labell{sstring}\eeqa
where the closed string vertex operator inserted at the middle and open string vertex
operators at the boundary of the disk world-sheet. 
The vertex operators in \reef{sstring} are given as\footnote{In
string theory, we set $\alpha'=2$.} \beqa
V_{A}^{(0)}(x) &=& \xi_{a}\bigg(\partial
X^a(x)+2ik\cd\psi\psi^a(x)\bigg)e^{2ik.X(x)},
\nonumber\\
V_{A}^{(-1)}(y) &=&\xi.\psi(y) e^{-\phi(y)} e^{2ik\cd X(y)},
\nonumber\\
V_{RR}^{(-\frac{1}{2},-\frac{1}{2})}(z,\bar{z})&=&(P_{-}\fsH_{(n)}M_p)^{\al\be}e^{-\phi(z)/2}
S_{\al}(z)e^{ip\cd X(z)}e^{-\phi(\bar{z})/2} S_{\be}(\bar{z})
e^{ip\cd D \cd X(\bar{z})},\nonumber\eeqa 
where $k$ is the momentum
of gauge field which satisfies the on-shell condition
$k^2=0$ and $k.\xi=0$.
The projector in the RR vertex operator is $P_{-} = \ha (1-\ga^{11})$  and
\begin{displaymath}
\fsH_{(n)} = \frac{a
_n}{n!}H_{\mu_{1}\ldots\mu_{n}}\ga^{\mu_{1}}\ldots
\ga^{\mu_{n}}
\ ,
\non\end{displaymath}
where $n=2,4$ for type IIA and $n=1,3,5$ for type IIB. $a_n=i$ for IIA and $a_n=1$ for IIB theory. The spinorial
indices are raised with the charge conjugation matrix, \ie
$(P_{-}\fsH_{(n)})^{\al\be} =
C^{\al\del}(P_{-}\fsH_{(n)})_{\del}{}^{\be}$ (for more conventions and
notations see appendix~B of~\cite{Garousi:1996ad}).  The
RR bosons are massless so $p^{2}=0$. The world-sheet  fields have been
extended to the entire complex plane. That is,  we have replaced 
\beqa
\frac{}{}
\nonumber\eeqa
\begin{displaymath}
\tilde{X}^{\mu}(\bar{z}) \rightarrow D^{\mu}_{\nu}X^{\nu}(\bar{z}) \ ,
\spa
\tilde{\psi}^{\mu}(\bar{z}) \rightarrow
D^{\mu}_{\nu}\psi^{\nu}(\bar{z}) \ ,
\spa
\tilde{\phi}(\bar{z}) \rightarrow \phi(\bar{z})\,, \mand
\tilde{S}_{\al}(\bar{z}) \rightarrow M_{\al}{}^{\be}{S}_{\be}(\bar{z})
 \ ,
\non\end{displaymath}
where
\begin{displaymath}
D = \left( \begin{array}{cc}
-1_{9-p} & 0 \\
0 & 1_{p+1}
\end{array}
\right) \ ,\,\, \mand 
M_p = \left\{\begin{array}{cc}\frac{\pm i}{(p+1)!}\ga^{a_{1}}\ga^{a_{2}}\ldots \ga^{a_{p+1}}
\eps_{a_{1}\ldots a_{p+1}}\,\,\,\,{\rm for\, p \,even}\\ \frac{\pm 1}{(p+1)!}\ga^{a_{1}}\ga^{a_{2}}\ldots \ga^{a_{p+1}}\ga_{11}
\eps_{a_{1}\ldots a_{p+1}} \,\,\,\,{\rm for\, p \,odd}\end{array}\right. 
\non\end{displaymath}
Using this doubling trick, one can find the standard holomorphic correlators for the world-sheet fields $X^{\mu},\psi^{\mu}, \phi$ as the following 
\begin{eqnarray}
\lan X^{\mu}(z)X^{\nu}(w)\ran & = & -\eta^{\mu\nu}\log(z-w) , \non \\
\lan \psi^{\mu}(z)\psi^{\nu}(w) \ran & = & -\eta^{\mu\nu}(z-w)^{-1} \ ,\non \\
\lan\phi(z)\phi(w)\ran & = & -\log(z-w) \ .
\labell{prop}\end{eqnarray}
Introducing $x_{4}\equiv\ z=x+iy$ and $x_{5}\equiv\bz=x-iy$,  the  amplitude  reduces to the following
correlators for 123 ordering
\beqa {\cal A}^{CAAA}&\sim& \int
 dx_{1}dx_{2}dx_{3}dx_{4} dx_{5}\,
(P_{-}\fsH_{(n)}M_p)^{\al\be}\xi_{1a}\xi_{2b}\xi_{3c}x_{45}^{-1/4}(x_{34}x_{35})^{-1/2}\nonumber\\&&
\times(I_1+I_2+I_3+I_4)\Tr(\lam_1\lam_2\lam_3),\labell{125}\eeqa where
$x_{ij}=x_i-x_j$, with the Wick theorem one can find the correlators as
\beqar
I_1&=&{<:\partial
X^a(x_1)e^{2ik_1.X(x_1)}:\partial X^b(x_2)e^{2ik_2.X(x_2)}
:e^{2ik_3.X(x_3)}:e^{ip.X(x_4)}:e^{ip.D.X(x_5)}:>}
 \  \non \\&&\times{<:S_{\al}(x_4):S_{\be}(x_5):\psi^c(x_3):>},\nonumber\\
I_2&=&{<:\partial X^a(x_1)e^{2ik_1.X(x_1)}:e^{2ik_2.X(x_2)}
:e^{2ik_3.X(x_3)}:e^{ip.X(x_4)}:e^{ip.D.X(x_5)}:>}
 \  \non \\&&\times{<:S_{\al}(x_4):S_{\be}(x_5):2ik_2.\psi\psi^{b}(x_2):\psi^c(x_3):>},\nonumber\\
 I_3&=&{<: e^{2ik_1.X(x_1)}:\partial X^b(x_2)e^{2ik_2.X(x_2)}
:e^{2ik_3.X(x_3)}:e^{ip.X(x_4)}:e^{ip.D.X(x_5)}:>}
 \  \non \\&&\times{<:S_{\al}(x_4):S_{\be}(x_5):2ik_1.\psi\psi^{a}(x_1):\psi^c(x_3):>},\nonumber\\
 I_4&=&{<: e^{2ik_1.X(x_1)}:e^{2ik_2.X(x_2)}
:e^{2ik_3.X(x_3)}:e^{ip.X(x_4)}:e^{ip.D.X(x_5)}:>}
 \  \non \\&&\times{<:S_{\al}(x_4):S_{\be}(x_5):2ik_{1}\cd\psi\psi^a(x_1):2ik_{2}\cd\psi\psi^b(x_2):\psi^c(x_3):>}.\eeqar
Using the first correlator,
 it is not difficult to calculate the correlators of $X$. We use the Wick-like rule~\cite{Liu:2001qa} 
and~\cite{Kostelecky:1986xg} to find the correlation function involving an arbitrary number of world-sheet fermions  ($\psi$s)
and two spin operators ($SS$).
One can generalize the Wick-like rule to find the correlation function of two spin operators and an arbitrary number of currents \cite{hatefi:2008}. 
 The only important point in using the Wick-like rule for currents is that one must not consider the Wick-like contraction for the two $\psi$s in one  current. 
 Taking this into account we can obtain the correlation function between two spin operators, one current and one world-sheet fermion as follows 
\beqa
I_5^{cbd}&=&<:S_{\al}(x_4):S_{\be}(x_5):\psi^d\psi^b(x_2):\psi^c(x_3):>\nonumber\\
&=&\bigg\{(\Gamma^{cbd}C^{-1})_{\alpha\beta}
+\frac{2Re[x_{24}x_{35}]}{x_{23}x_{45}}\bigg(\eta^{dc}(\gamma^{b}C^{-1})_{\alpha\beta}-\eta^{bc}(\gamma^{d}C^{-1})_{\alpha\beta}\bigg)\bigg\}
\nonumber\\&&\times2^{-3/2}x_{45}^{1/4}(x_{24}x_{25})^{-1}(x_{34}x_{35})^{-1/2}. 
\label{68}\eeqa
The calculation of the correlation function between two spin operators, two currents and one world-sheet fermion is more complicated,
 but using this generalization it is simply given by 
\beqa
I_6^{cbeaf}&=&<:S_{\al}(x_4):S_{\be}(x_5):\psi^f\psi^a(x_1):\psi^e\psi^b(x_2):\psi^c(x_3):>\nonumber\\
&=&\bigg\{(\Gamma^{cbeaf}C^{-1})_{{\alpha\beta}}+2r_1\frac{Re[x_{14}x_{25}]}{x_{12}x_{45}}+2r_2\frac{Re[x_{14}x_{35}]}{x_{13}x_{45}}+2r_3\frac{Re[x_{24}x_{35}]}{x_{23}x_{45}}+4r_4\nonumber\\&&\times\bigg(\frac{Re[x_{14}x_{25}]}{x_{12}x_{45}}\bigg)^{2}
+4r_5\bigg(\frac{Re[x_{14}x_{25}]}{x_{12}x_{45}}\times\frac{Re[x_{14}x_{35}]}{x_{13}x_{45}}\bigg)+4r_6\bigg(\frac{Re[x_{14}x_{25}]}{x_{12}x_{45}}\nonumber\\&&\times\frac{Re[x_{24}x_{35}]}{x_{23}x_{45}}\bigg)
\bigg\}2^{-5/2}x_{45}^{5/4}(x_{14}x_{15}x_{24}x_{25})^{-1}(x_{34}x_{35})^{-1/2},\label{hh}\eeqa
where 
\beqa
r_1&=&\bigg(\eta^{fe}(\Gamma^{cba}C^{-1})_{\alpha\beta}
-\eta^{fb}(\Gamma^{cea}C^{-1})_{\alpha\beta}-\eta^{ae}(\Gamma^{cbf}C^{-1})_{\alpha\beta}+\eta^{ab}(\Gamma^{cef}C^{-1})_{\alpha\beta}\bigg),\nonumber\\
r_2&=&\bigg(\eta^{fc}(\Gamma^{bea}C^{-1})_{\alpha\beta}
-\eta^{ac}(\Gamma^{bef}C^{-1})_{\alpha\beta}\bigg),\nonumber\\
r_3&=&\bigg(\eta^{ec}(\Gamma^{baf}C^{-1})_{\alpha\beta}
-\eta^{bc}(\Gamma^{eaf}C^{-1})_{\alpha\beta}\bigg),\nonumber\\
r_4&=&\bigg((-\eta^{fe}\eta^{ab}+\eta^{fb}\eta^{ae})(\gamma^{c}C^{-1})_{\alpha\beta}\bigg),\nonumber\\
r_5&=&\bigg((\eta^{fe}\eta^{ac}-\eta^{fc}\eta^{ae})(\gamma^{b}C^{-1})_{\alpha\beta}+(-\eta^{fb}\eta^{ac}+\eta^{fc}\eta^{ab})(\gamma^{e}C^{-1})_{\alpha\beta}\bigg),\nonumber\\
r_6&=&\bigg((-\eta^{fe}\eta^{bc}+\eta^{fb}\eta^{ec})(\gamma^{a}C^{-1})_{\alpha\beta}+(\eta^{ae}\eta^{bc}-\eta^{ab}\eta^{ec})(\gamma^{f}C^{-1})_{\alpha\beta}\bigg).
\eeqa
Replacing the above spin correlators in \reef{125} and performing the correlators over $X$, one finds:
\beqa
{\cal A}^{CAAA}&\!\!\!\!\sim\!\!\!\!\!&\int dx_{1}dx_{2} dx_{3}dx_{4}dx_{5}(P_{-}\fsH_{(n)}M_p)^{\al\be}I\xi_{1a}\xi_{2b}\xi_{3c}x_{45}^{-1/4}(x_{34}x_{35})^{-1/2}\nonumber\\&&\times
\bigg(I_7^c(-\eta^{ab}x_{12}^{-2}+a^a_1a^b_2)+a^a_1a^{cb}_3+a^b_2a^{ca}_4-4k_{1f}k_{2e}I_6^{cbeaf}\bigg)\Tr(\lam_1\lam_2\lam_3)\labell{amp3},\eeqa
where  $I_6^{cbeaf}$ is given in \reef{hh} and
\beqa
I&=&|x_{12}|^{4k_1.k_2}|x_{13}|^{4k_1.k_3}|x_{14}x_{15}|^{2k_1.p}|x_{23}|^{4k_2.k_3}|x_{24}x_{25}|^{2k_2.p}
|x_{34}x_{35}|^{2k_3.p}|x_{45}|^{p.D.p},\nonumber\\
a^a_1&=&-ik_2^{a}\bigg(\frac{x_{42}}{x_{41}x_{12}}+\frac{x_{52}}{x_{51}x_{12}}\bigg)
-ik_3^{a}\bigg(\frac{x_{43}}{x_{41}x_{13}}+\frac{x_{53}}{x_{51}x_{13}}\bigg),\nonumber\\
a^b_2&=&-ik_1^{b}\bigg(\frac{x_{14}}{x_{42}x_{12}}+\frac{x_{15}}{x_{52}x_{12}}\bigg)
-ik_3^{b}\bigg(\frac{x_{43}}{x_{42}x_{23}}+\frac{x_{53}}{x_{52}x_{23}}\bigg),\nonumber\\
a^{cb}_3&=&2ik_{2d}I_5^{cbd},\nonumber\\
a^{ca}_4&=&2ik_{1e}2^{-3/2}x_{45}^{1/4}(x_{14}x_{15})^{-1}(x_{34}x_{35})^{-1/2} \nonumber\\&&\times\bigg\{(\Gamma^{cae}C^{-1})_{\alpha\beta}
+\frac{2Re[x_{14}x_{35}]}{x_{13}x_{45}}\bigg(\eta^{ce}(\gamma^{a}C^{-1})_{\alpha\beta}-\eta^{ac}(\gamma^{e}C^{-1})_{\alpha\beta}\bigg)\bigg\}
,\nonumber\\
I_7^c&=&<:S_{\al}(x_4):S_{\be}(x_5):\psi^c(x_3):>=2^{-1/2}x_{45}^{-3/4}(x_{34}x_{35})^{-1/2}
(\gamma^{c}C^{-1})_{\alpha\beta}.\nonumber\eeqa
One can show that the integrand is invariant under
SL(2,R) transformation. Gauge fixing this symmetry by fixing the position of the open string vertex operators as  \beqar
 x_{1}=0 ,\qquad x_{2}=1,\qquad x_{3}\rightarrow \infty,
 \qquad dx_1dx_2dx_3\rightarrow x_3^{2}.
 \eeqar
 one finds the following integral
\beqa
 \int d^2 \!z |1-z|^{a} |z|^{b} (z - \bar{z})^{c}
(z + \bar{z})^{d},
 \eeqa
 where $d=0,1,2$ and $a,b,c$ are given in terms of the Mandelstam variables,\ie
\beqar
s&=&-(k_1+k_3)^2,\qquad t=-(k_1+k_2)^2,\qquad u=-(k_2+k_3)^2.
\qquad\eeqar
The region of integration  is the upper half of the complex plane. For $d=0,1$ the result is given in \cite{Fotopoulos:2001pt}
and for $d=2$ the result is given in \cite{hatefi:2008}.
Using those integrals one can write the amplitude \reef{amp3} as
\beqa {\cal A}^{AAAC}&=&{\cal A}_{1}+{\cal A}_{2}+{\cal A}_{3},\labell{11u}\eeqa
where
\beqa
{\cal A}_{1}&\!\!\!\sim\!\!\!&2^{1/2}\xi_{1a}\xi_{2b}\xi_{3d}k_{1f}k_{2e}
\Tr(P_{-}\fsH_{(n)}M_p\Gamma^{dbeaf}
)(t+s+u)L_1, 
\nonumber\\
{\cal A}_{2}&\sim&2^{-1/2}\Tr(P_{-}\fsH_{(n)}M_p \Gamma^{dba})\bigg\{-tL_2\xi_{1a}\xi_{2b}\xi_{3d}+\bigg[L_2(-2k_1.\xi_2k_{2b}
\xi_{1a}\xi_{3d}-2k_2.\xi_1k_{1a}\xi_{2b}\xi_{3d}\nonumber\\&&-2k_2.\xi_1k_{2a}\xi_{2b}\xi_{3d}
+2k_1.\xi_2k_{1a}\xi_{1b}\xi_{3d}+2\xi_{1}.\xi_{2}k_{1a}k_{2b}\xi_{3d})\bigg]
-\bigg[3\leftrightarrow 1\bigg]-\bigg[3\leftrightarrow 2\bigg]
\bigg\},
\nonumber\\
{\cal A}_{3}&\sim&2^{-1/2}L_1\bigg\{\bigg[\Tr(P_{-}\fsH_{(n)}M_p\xi_3.\gamma)
(-2tk_3.\xi_1k_3.\xi_2+2uk_3.\xi_1k_1.\xi_2
+2sk_2.\xi_1k_3.\xi_2\nonumber\\&&+us\xi_1.\xi_2)\bigg]+\bigg[3\leftrightarrow 2\bigg]+\bigg[3\leftrightarrow 1\bigg]+\bigg(\bigg[
\Tr(P_{-}\fsH_{(n)}M_pk_2.\gamma)(-2tk_3.\xi_1\xi_2.\xi_3\nonumber\\&&+2uk_1.\xi_2\xi_1.\xi_3+2sk_2.\xi_1\xi_3.\xi_2-2u\xi_1.\xi_2k_1.\xi_3)\bigg]
+\bigg[2\leftrightarrow 1\bigg]\bigg)\bigg\}.
\labell{48}\eeqa
where the functions
 $L_1,L_2$ are the following 
\beqa
L_1&=&(2)^{-2(t+s+u)}\pi{\frac{\Gamma(-u+\frac{1}{2})
\Gamma(-s+\frac{1}{2})\Gamma(-t+\frac{1}{2})\Gamma(-t-s-u)}
{\Gamma(-u-t+1)\Gamma(-t-s+1)\Gamma(-s-u+1)}},\nonumber\\
L_2&=&(2)^{-2(t+s+u)}\pi{\frac{\Gamma(-u+1)
\Gamma(-s+1)\Gamma(-t)\Gamma(-t-s-u+\frac{1}{2})}
{\Gamma(-u-t+1)\Gamma(-t-s+1)\Gamma(-s-u+1)}}.\nonumber\eeqa
Since $\fsH_{(n)}, M_p, \Gamma^{dbeaf}$ and $\Gamma^{dea}$ are totally antisymmetric combinations of the
Gamma matrices, one can then understand that the amplitude is non zero for $p=n+4,  p=n+2$ and $p=n$. From the poles of the gamma functions, one can easily see that the scattering amplitude has infinite massless poles and  infinite number of massive poles. To compare this with the field theory, which has massless fields, one must expand the amplitude such that the  massless poles of the field theory survive  and all other poles vanish in the form of contact terms. In the next section we use the low energy limit expansion by sending all Mandelstam variables to zero.
\section{Momentum expansion}
We want to examine the limit of $\alpha'\rightarrow 0$  of the above string amplitude. Using the  momentum conservation along the world volume of brane,  $k_1^{a} + k_2^{a}+k_3^{a}+p^{a} =0$, one finds the Mandelstam variables satisfy
the following constraint 
\beqa
s+t+u=-p_ap^a.
\labell{cons}\eeqa
 It has been argued in \cite{Garousi:2007si}, generally speaking that the momentum expansion of a S-matrix element should be around  $(k_i+k_j)^2\rightarrow 0$ and/or $k_i\inn k_j\rightarrow 0$.  The case $(k_i+k_j)^2\rightarrow 0$ is when there is a massless pole in the $(k_i+k_j)^2$-channel. Notice that the amplitude \reef{sstring} must have only  massless poles in the $(k_1+k_2)^2$, $(k_1+k_3)^2$ and $(k_2+k_3)^2$-channels, so correct momentum expansion at the low energy limit for t-channel must be around
 \beqa
(k_1+k_2)^2\rightarrow 0,\qquad k_1.k_3\rightarrow 0,\qquad k_2.k_3\rightarrow 0. \nonumber\eeqa
Also the correct momentum expansion for s,u-channels respectively are
\beqa
(k_1+k_3)^2\rightarrow 0,\qquad k_1.k_2\rightarrow 0,\qquad k_2.k_3\rightarrow 0, \nonumber\\
(k_2+k_3)^2\rightarrow 0,\qquad k_1.k_2\rightarrow 0,\qquad k_1.k_3\rightarrow 0, \nonumber\eeqa
Using the on-shell relations $k_1^2=k_2^2=k_3^2=0$ one can rewrite them in terms of the Mandelstam variables as
\beqa
s\rightarrow 0,\qquad t\rightarrow 0,\qquad  u\rightarrow 0. \labell{point}\eeqa
Including the constraint \reef{cons}, one should realize that $p_ap^a\rightarrow 0$ which is possible for D-branes. Therefore the S-matrix
element can be evaluated for BPS branes. 

Expansion of the functions $L_1,L_2$
around the above point is
\beqa
L_1&=&-{\pi^{5/2}}\left( \sum_{n=0}^{\infty}c_n(s+t+u)^n\right.
\left.+\frac{\sum_{n,m=0}^{\infty}c_{n,m}[s^nu^m +s^mu^n]}{(t+s+u)}\right.\nonumber\\
&&\left.+\sum_{p,n,m=0}^{\infty}f_{p,n,m}(s+t+u)^p[(s+u)^{n}(su)^{m}]\right),\nonumber\\
L_2&=&-\pi^{3/2}\bigg(\frac{1}{t}\sum_{n=-1}^{\infty}b_n(u+s)^{n+1}+\sum_{p,n,m=0}^{\infty}e_{p,n,m}t^{p}(su)^{n}(s+u)^m\bigg).\labell{high}\eeqa
where some of the coefficients $b_n,\,e_{p,n,m},\,c_n,\,c_{n,m}$ and $f_{p,n,m}$ are
\beqa 
&&b_{-1}=1,\,b_0=0,\,b_1=\frac{1}{6}\pi^2,\,b_2=2\z(3),c_0=0,c_1=-\frac{\pi^2}{6},\nonumber\\
&&e_{2,0,0}=e_{0,1,0}=2\z(3),e_{1,0,0}=\frac{1}{6}\pi^2,e_{1,0,2}=\frac{19}{60}\pi^4,e_{1,0,1}=e_{0,0,2}=6\z(3),\nonumber\\
&&e_{0,0,1}=\frac{1}{3}\pi^2,e_{3,0,0}=\frac{19}{360}\pi^4,e_{0,0,3}=e_{2,0,1}=\frac{19}{90}\pi^4,e_{1,1,0}=e_{0,1,1}=\frac{1}{30}\pi^4,\labell{577}\\
&&c_2=-2\xi(3),
\,c_{1,1}=\frac{\pi^2}{6},\,c_{0,0}=\frac{1}{2},c_{3,1}=c_{1,3}=\frac{2}{15}\pi^4,c_{2,2}=\frac{1}{5}\pi^4,\nonumber\\
&&c_{1,0}=c_{0,1}=0,
c_{3,0}=c_{0,3}=0\, 
,\,c_{2,0}=c_{0,2}=\frac{\pi^2}{6},c_{1,2}=c_{2,1}=-4\xi(3),\nonumber\\
&&f_{0,1,0}=\frac{\pi^2}{3},\,f_{0,2,0}=-f_{1,1,0}=-6\xi(3),f_{0,0,1}=-2\xi(3),c_{4,0}=c_{0,4}=\frac{1}{15}\pi^4.\, \nonumber
\eeqa
Note that the coefficients $b_n$ are exactly the coefficients that appear in the momentum expansion of the S-matrix element of one RR, two gauge fields and one tachyon vertex operator \cite{hatefi:2008}.
Meanwhile $c_n,\,c_{n,m},f_{p,n,m}$ are different from those coefficients which appeared in \cite {hatefi:2008}. The function of
$L_1$ has infinite massless poles in the $(t+s+u)$-channel and $L_2$ has infinite massless poles in the $t$-channel. These poles must  be reproduced in field theory by appropriate couplings. Let us study each case separately.
\section{Low energy field theory }
We are
interested in the part of effective field theory of
D-branes which includes only gauge fields. It should be possible to extract the necessary terms from the covariant Born-Infeld action
constructed as the effective D-brane action in \reef{BI} .
The Born-Infeld action is an action for all orders of $\alpha'$ (see for more details \cite{an,esf}). The low energy non-abelian extension of the action was proposed to be
the symmetrized trace of non-abelian generalization of Born-Infeld action (with flat background in the bulk) \cite{aat}.
There, it was shown that defining
non abelian Born-Infeld action with this trace produced the known results for the
scattering of gauge fields up to fourth order in the field strengths \cite{dg}. However, there are some reasons which indicate
the symmetrized trace prescription does not work for $F^6$ \cite{ah}. Then, it was proved
that the symmetrized trace requires corrections at sixth order \cite{Hashimoto:1997gm}. 
Using noncommutative  field theory some efforts for the form of BI action were done \cite{Cornalba:1999ah}.

The non-abelian field strength and covariant
derivative of the gauge field are defined respectively as
 \beqa F^{ab}=
\partial^aA^b-\partial^bA^a-i[A^a,A^b],~~~~
D_aF_{bc}=\partial_aF_{bc}-i[A_a,F_{bc}]. \nonumber\eeqa
where $A_a=A_a^{\alpha}\Lambda_{\alpha}$
and $\Lambda_{\alpha}$ are the hermitian matrices.
 Our conventions for $\Lambda^{\alpha}$ are
\beqa
\sum_\alpha
\Lambda^{\alpha}_{ij}\Lambda^{\alpha}_{kl}=\delta_{ik}\delta_{jl}\,\,,\,\,~~~
\Tr(\Lambda^{\alpha}\Lambda^{\beta})&\!\!\!=\!\!\!&\delta^{\alpha\beta}.
\nonumber\eeqa
Using the following expression, one can expand the square root in
the non-abelian action \reef{BI} to  produce various interacting
terms \cite{BitaghsirFadafan:2006cj} :
\beqa 
\sqrt{ -\det(M_0+M)}&=&\sqrt{
-\det(M_0)}\bigg(1+\frac{1}{2}Tr(M_{0}^{-1}M)-\frac{1}{4}Tr(M_{0}^{-1}MM_{0}^{-1}M)
\nonumber\\&&+\frac{1}{6}Tr(M_{0}^{-1}MM_{0}^{-1}MM_{0}^{-1}M)
-\frac{1}{8}Tr(M_{0}^{-1}MM_{0}^{-1}MM_{0}^{-1}MM_{0}^{-1}M)\nonumber\\&&
+\frac{1}{8}(Tr(M_{0}^{-1}M))^2-\frac{1}{8}(Tr(M_{0}^{-1}M))^3+\frac{1}{32}(Tr(M_{0}^{-1}MM_{0}^{-1}M))^2\nonumber\\&&
+\cdots\bigg).
\labell{q13}\eeqa
In \reef{BI},  $M_0$ and $M$ are \beqa
M_0&=&\eta_{ab},\nonumber\\M&=&2\pi\alpha' F_{ab}.
\nonumber\eeqa 
The terms of the above expansion which have
contribution to the S-matrix element \reef{sstring} are in the following  
\beqa {\cal
L}&=&-T_p(\pi\alpha')\Tr\left(-(\pi\alpha')F_{ab}F^{ba}\right)\labell{expandL}\\
&&-T_p(2\pi\alpha')^4S\Tr\left(-\frac{1}{8}F_{bd}F^{df}F_{fh}F^{hb}+\frac{1}{32}(F_{ab}F^{ba})^2\right).
\nonumber\eeqa
 Note that after averaging all
possible permutations of the above terms in the second line  \reef{expandL}, one must take overall trace over the group theory indices.
 The couplings in the second line have been confirmed \cite {Kitazawa:1987xj}. 
 We want to obtain the higher derivative couplings of the four gauge fields  
 and then show that 
 these terms reproduce infinite massless poles in the S-matrix.

\subsection{$p=n+4$ case}

 This  is the simplest case to consider. Only  ${\cal A}_1$ in \reef{48} is non-zero. One can calculate the trace of ${\cal A}_1$ as follows 
\beqa
\Tr\bigg(P_{-}\fsH_{(n)}M_p\Gamma^{dbeaf}\bigg)&=&\pm\frac{32}{2n!}\eps^{dbeafa_{0}\cdots a_{p-5}}H_{a_{0}\cdots a_{p-5}},\nonumber\eeqa
 We are going to compare string theory S-matrix elements with field theory including their coefficients, however  we are not interested in fixing the overall sign of the amplitudes.
Taking into account the above trace, the string amplitude becomes
\beqa
{\cal A}^{CAAA}=\pm\frac{32}{(p-4)!}\mu_p \pi^{1/2}\Tr(\lam_1\lam_2\lam_3)\xi_{1a}\xi_{2b}\xi_{3d}k_{1f}k_{2e}
\eps^{dbeafa_{0}\cdots a_{p-5}}H_{a_{0}\cdots a_{p-5}}(s+t+u)L_1,\label{111}
\eeqa
where we normalized the amplitude by $ (\mu_p 2^{1/2}\pi^{1/2})$.
The above amplitude is antisymmetric upon interchanging the gauge fields. So the whole amplitude is zero for an abelian gauge group. The amplitude also satisfies the Ward identity, \ie it vanishes under replacing  each of  $\xi^i\rightarrow k^i$. Since  $(t+s+u)L_1$ has no tachyon/massless pole, then the amplitude  has only contact terms. The leading contact term is reproduced by the following coupling 
\beqa
\frac{1}{3!}\mu_p(2\pi\alpha')^{3}\Tr (C_{p-5}\wedge F\wedge F\wedge F).\labell{hderv}
\eeqa
The non-leading order terms should correspond to the higher derivative extension of the coupling. 
\subsection{$p=n$ case}

The next simple case is  $p=n$. Only ${\cal A}_3$ in \reef{48} is non-zero  for this case. The calculation of the trace in this 
part of the amplitude is
\beqa
\Tr\bigg(\fsH_{(n)}M_p\gamma^{a}
\bigg)&=&\pm\frac{32}{n!}\eps^{a_{0}\cdots a_{p-1}a}H_{a_{0}\cdots a_{p-1}},
\nonumber\eeqa
Substituting  this trace in  ${\cal A}_3$,  one finds
\beqa
{\cal A}^{CAAA}&=&\pm\frac{32}{2p!}\mu_p\pi^{1/2}L_1\Tr(\lam_1\lam_2\lam_3)\eps^{a_{0}\cdots a_{p-1}a}H_{a_{0}\cdots a_{p-1}}\bigg\{\bigg[
\xi_{3a}(-2tk_3.\xi_1k_3.\xi_2+2uk_3.\xi_1\nonumber\\&&\times k_1.\xi_2
+2sk_2.\xi_1k_3.\xi_2+us\xi_1.\xi_2)\bigg]+\bigg[3\leftrightarrow 2\bigg]
+\bigg[3\leftrightarrow 1\bigg]+\bigg(\bigg[
k_{2a}(-2tk_3.\xi_1\xi_2.\xi_3\nonumber\\&&+2uk_1.\xi_2\xi_1.\xi_3+2sk_2.\xi_1\xi_3.\xi_2-2u\xi_1.\xi_2k_1.\xi_3)\bigg]
+\bigg[2\leftrightarrow 1\bigg]\bigg)\bigg\}.\labell{47}\eeqa
A check of our calculations is that the above amplitude  satisfies the Ward identity associated
with the gauge invariance of the open string states.
 The amplitude is symmetric under interchanging $1\leftrightarrow 2$. So the amplitude is non-zero even for the abelian case.
 All terms in \reef{47} have infinite massless poles in the $(s+t+u)$-channel and infinite contact terms. In the next section, firstly we want to reproduce the first massless pole using the symmetric trace prescription of BI action. Then we find higher derivative couplings of four gauge fields in order to show that massless poles can be produced to all orders of $\alpha'$ by WZ coupling $C_{p}\wedge F$ and by the higher derivative couplings of four gauge fields.

\subsection{First massless pole for p=n case}

 The terms in the second line of \reef{expandL} give
four gauge field couplings.  In order to reproduce the first massless pole from the couplings \reef{expandL} we should consider figure 1 as the
 Feynman diagram for $p=n$ case.
\begin{center}
\begin{picture}
(600,125)(0,0)
\Photon(125,105)(185,70){4}{7.5}\Text(150,105)[]{$A_{1}$}
\Photon(125,35)(185,70){4}{7.5}\Text(150,39)[]{$A_{3}$}
\Photon(125,70)(185,70){4}{7.5}\Text(145,79)[]{$A_2$}
\Photon(185,70)(255,70){4}{7.5}\Text(220,88)[]{$A$}
\Gluon(255,70)(295,105){4.20}{5}\Text(275,105)[]{$C_{p-1}$}
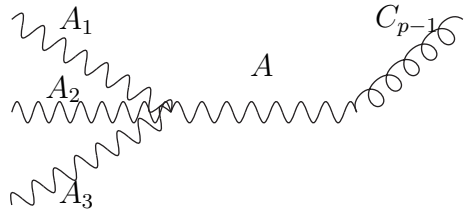
\end{picture}\\ {\sl Figure 1 : The Feynman diagram corresponding to the amplitudes \reef{47}.}  
\end{center}
Since the propagator is abelian, we must calculate three possible permutations 
to obtain the desired 123 ordering. Writing symmetric traces in terms of ordinary traces (apart from overall factor),  one can write the two last terms in \reef{expandL} as $(L_5^{0,0}+L_6^{0,0}+L_7^{0,0})$ which
 are the following :
\beqa
L_5^{0,0}&=&-\frac{1}{4\pi^2}\Tr\bigg(a_{0,0}(F_{bd}F^{df}F_{fh}F^{hb})+b_{0,0}(F_{bd}F_{fh}F^{df}F^{hb})\bigg),\nonumber\\
L_6^{0,0}&=&-\frac{1}{4\pi^2}\Tr\bigg(a_{0,0}(F_{bd}F^{df}F^{hb}F_{fh})+b_{0,0}(F_{bd}F^{hb}F^{df}F_{fh})\bigg),\nonumber\\
L_7^{0,0}&=&\frac{1}{8\pi^2}\Tr\bigg(a_{0,0}(F_{ab}F^{ab}F_{cd}F^{cd})+b_{0,0}(F_{ab}F^{cd}F^{ab}F_{cd})\bigg).\label{25}\eeqa
where $a_{0,0}=\frac{-\pi^2}{6},b_{0,0}=\frac{-\pi^2}{12}$.

The massless poles of the amplitude \reef{47} are given by the following amplitude 
\beqa
{\cal A}&=&V_{\alpha}^{a}(C_{p-1},A)G_{\alpha\beta}^{ab}(A)V_{\beta}^{b}(A,A_1,
A_2,A_3),\labell{amp549}\eeqa
where the gauge field propagator and the vertex $V_{\alpha}^{a}(C_{p-1},A)$ are given as
\beqa
G_{\alpha\beta}^{ab}(A) &=&\frac{i\delta_{\alpha\beta}\delta^{ab}}{(2\pi\alpha')^2 T_p
(s+t+u)},\nonumber\\
V_{\alpha}^{a}(C_{p-1},A)&=&i(2\pi\alpha')\mu_p\frac{1}{p!}\epsilon^{a_0\cdots a_{p-1}a}H_{a_0\cdots a_{p-1}}\Tr(\Lambda_{\alpha}).
\labell{Fey}
\eeqa
where $V_{\alpha}^{a}(C_{p-1},A)$  has been found in \cite{Garousi:2007fk}. In the above vertex $\Tr(\Lambda_{\alpha})$ is non-zero for the abelian matrix $\Lambda_{\alpha}$. The vertex $ V_{\beta}^{b}(A,A_1,
A_2,A_3)$  can be obtained from the four gauge field couplings of \reef{25}  as follows 
\beqa
I_8&=&V_{\beta}^{b}(A,A_1,A_2,A_3)=(2\pi\alpha')^4T_{p}\frac{1}{4}\Tr(\lam_1\lam_2\lam_3\Lambda_{\beta})\bigg\{\xi_{3}^{b}\bigg[
2tk_3.\xi_1k_3.\xi_2-2uk_3.\xi_1k_1.\xi_2\nonumber\\&&
-2sk_2.\xi_1k_3.\xi_2-us\xi_1.\xi_2\bigg]+\bigg[3\leftrightarrow 2\bigg]+\bigg[3\leftrightarrow 1\bigg]
+\bigg(k_{2}^{b}\bigg[
+2tk_3.\xi_1\xi_2.\xi_3-2uk_1.\xi_2\xi_1.\xi_3\nonumber\\&&+2uk_1.\xi_3\xi_1.\xi_2-2s\xi_2.\xi_3k_2.\xi_1\bigg]+\bigg[2\leftrightarrow 1\bigg]\bigg)
\bigg\},\labell{amp56}\eeqa
where $k_1,k_2,k_3$  are the momenta of on-shell gauge fields. 
Replacing the above vertex in \reef{amp549} we find
 exactly the first massless pole of the equation of \reef{47}. 
In order to obtain all infinite massless poles of the amplitude for the $p=n$ case we should find higher derivative couplings of four gauge fields. 
\section{Four gauge field couplings }

The S-matrix element of all four point massless vertex operators in superstring theory was calculated in standard books \cite{mgjs,jp}. For two important reasons   one can find  the higher derivative couplings of four gauge fields from higher derivative couplings of four scalar fields \cite{Garousi:2008xp}
by using T-duality transformation.

The first reason is that Mandelstam Variables for both four gauge fields amplitude and four massless scalar fields amplitude satisfy the constraint of $s+t+u=0$. Also the massless poles of the Feynman amplitude resulting from the non abelian kinetic term of the scalars and gauge fields are reproduced at the low energy limit by sending $s,t,u\rightarrow 0$. The second reason is that, external states in both of them satisfy the on-shell condition $k_i^2=0 $ and physical state condition $k.\xi=0$ (in fact, the behavior of massless transverse scalars is similar to world volume gauge fields). Also both of them transform in the adjoint representation
of U(N) group. 

One may expect that the higher derivative couplings of four gauge fields should be similar to the higher derivative couplings of four scalar fields.
The only difference is related to their polarization. Gauge fields' polarization only has components in the world volume direction while
scalar fields' polarization has transverse components on the D-brane ,$\ie$ the physical state condition for gauge field is $k_1.\xi_1=k_2.\xi_2=\cdots=k_n.\xi_n=0$ 
while for scalar fields it satisfies $k_1.\xi_1=k_1.\xi_2=\cdots=k_i.\xi_j=0 $ where $i,j=1,2,...n$.

On the other hand the S-matrix element of the scalar field vertex operators can be read from the S-matrix element of gauge
field vertex operators by restricting the polarization of the gauge fields to transverse directions and the
momentum of the gauge fields to the world-volume directions. To find four gauge field couplings to all orders of $\alpha'$, we follow the steps mentioned in \cite{Garousi:2008xp}.
The massless poles in \reef{47} are reproduced by the non-abelian kinetic terms of the gauge field and the contact terms with coefficients $a_{0,0}$ and $b_{0,0}$ are also reproduced (apart from an over all factor) by the following terms 
\beqa
-T_pS\Tr\left(-\frac{1}{8}F_{bd}F^{df}F_{fh}F^{hb}+\frac{1}{32}(F_{ab}F^{ba})^2\right).
\label{55}\eeqa 
Meanwhile for the scalar field they are reproduced by 
\beqa
&&- T_p{\rm STr}
\left(-\frac{1}{4}D_a\phi^iD_b\phi_iD^b\phi^jD^a\phi_j+\frac{1}{8}
(D_a\phi^i D^a\phi_i)^2\right).\labell{a011}\eeqa 
Note that the differences between \reef{55} and \reef{a011} are the coefficients and indices. Therefore using T-duality transformation
not only should one substitute the covariant derivative of the scalar field $D\phi$ into the field strength of the gauge field $F$, but also replace
transverse indices with world volume indices to find higher derivative couplings of four gauge fields from higher derivative couplings of four scalar fields (the equation (35) of \cite{Garousi:2008xp}). Performing symmetric traces in terms of ordinary traces one can write \reef{55} as \reef{25}.
 Now one can extend it to the higher derivative terms as
\beqa
(2\pi\alpha')^4\frac{1}{8\pi^2}T_p\left(\alpha'\right)^{n+m}\sum_{m,n=0}^{\infty}(\cL_{5}^{nm}+\cL_{6}^{nm}+\cL_{7}^{nm}),\labell{highder}\eeqa
with
\beqa
&&\cL_{5}^{nm}=-
\Tr\left(\frac{}{}a_{n,m}\cD_{nm}[F_{bd}F^{df}F_{fh}F^{hb}]+\frac{}{} b_{n,m}\cD'_{nm}[F_{bd}F_{fh}F^{df}F^{hb}]+h.c.\frac{}{}\right),\nonumber\\
&&\cL_{6}^{nm}=-\Tr\left(\frac{}{}a_{n,m}\cD_{nm}[F_{bd}F^{df}F^{hb}F_{fh}]+\frac{}{}b_{n,m}\cD'_{nm}[F_{bd}F^{hb}F^{df}F_{fh}]+h.c.\frac{}{}\right),\nonumber\\
&&\cL_{7}^{nm}=\frac{1}{2}\Tr\left(\frac{}{}a_{n,m}\cD_{nm}[F_{ab}F^{ab}F_{cd}F^{cd}]+\frac{}{}b_{n,m}\cD'_{nm}[F_{ab}F^{cd}F^{ab}F_{cd}]+h.c\frac{}{}\right),\nonumber\eeqa
where the higher derivative operators 
$D_{nm} $ and $ D'_{nm}$ are defined \cite{Garousi:2008xp} as
\beqa
\cD_{nm}(EFGH)&\equiv&D_{b_1}\cdots D_{b_m}D_{a_1}\cdots D_{a_n}E  F D^{a_1}\cdots D^{a_n}GD^{b_1}\cdots D^{b_m}H,\nonumber\\
\cD'_{nm}(EFGH)&\equiv&D_{b_1}\cdots D_{b_m}D_{a_1}\cdots D_{a_n}E   D^{a_1}\cdots D^{a_n}F G D^{b_1}\cdots D^{b_m}H.\nonumber\eeqa
Of course the above couplings are exact up to total derivative terms and terms like $\prt_a\prt^aFFFF$ which are zero on-shell.
 Also these terms have no effect on the massless poles of S-matrix elements because by
canceling $k^2$ with the massless propagator one finds a contact term. These are the higher derivative extensions of four gauge field couplings of the action \reef{BI}.

\subsection{Infinite massless poles for $p=n$ case }
Here we would like to check that the infinite four gauge field couplings \reef{highder} produce infinite massless poles of the string theory S-matrix element \reef{47} which are in the $(s+t+u)$-channel. In fact they can be reproduced by WZ coupling $C_{p}\wedge F$ and by the higher derivative four gauge field couplings that have been found in \reef{highder}. For this aim, consider the amplitude of the decay of one R-R field to three gauge fields in the world-volume theory of the BPS branes which is given by the Feynman amplitude  
\reef{amp549}, where the gauge field propagator and the vertex $V_{\alpha}^{a}(C_{p-1},A)$ are in \reef{Fey}.
According to the fact that the off-shell gauge field must be abelian, one finds the higher derivative vertex  $ V_{\beta}^b(A,A_1,A_2,A_3)$  from the higher derivative couplings in \reef{highder} to be 
\beqa
&&\frac{I_8}{2\pi^2}(\alpha')^{n+m}(a_{n,m}+b_{n,m})
\bigg(\frac{}{}(k_3\inn k_1)^n(k_1\inn k)^m+(k_1\inn k)^m(k_2\inn k)^n 
+(k_1\inn k)^n(k_1\inn k_3)^m\nonumber\\&&+(k_1\inn k_3)^m (k_3\inn k_2)^n
+(k_3\inn k_1)^n(k_2\inn k_3)^m+(k_3\inn k_2)^m(k_2\inn k)^n
+(k_1\inn k)^n(k_2\inn k)^m\nonumber\\&&+(k_3\inn k_2)^n(k_2\inn k)^m\bigg),\labell{veraatt}\eeqa
where $I_8$ is in \reef{amp56} and $k$ is the momentum of the off-shell gauge field. Note that we must consider all 12 possible cyclic permutations to obtain the desired 123 ordering
of the amplitude. Some of the coefficients $a_{n,m}$ and $b_{n,m}$ are \cite{Garousi:2008xp}
\beqa
&&a_{0,0}=-\frac{\pi^2}{6},\,b_{0,0}=-\frac{\pi^2}{12},a_{1,0}=2\z(3),\,a_{0,1}=0,\,b_{0,1}=-\z(3),a_{1,1}=a_{0,2}=-7\pi^4/90,\nonumber\\
&&a_{2,2}=(-83\pi^6-7560\z(3)^2)/945,b_{2,2}=-(23\pi^6-15120\z(3)^2)/1890,a_{1,3}=-62\pi^6/945,\nonumber\\
&&\,a_{2,0}=-4\pi^4/90,\,b_{1,1}=-\pi^4/180,\,b_{0,2}=-\pi^4/45,a_{0,4}=-31\pi^6/945,a_{4,0}=-16\pi^6/945,\nonumber\\
&&a_{1,2}=a_{2,1}=8\z(5)+4\pi^2\z(3)/3,\,a_{0,3}=0,\,a_{3,0}=8\z(5),b_{1,3}=-(12\pi^6-7560\z(3)^2)/1890,\nonumber\\
&&a_{3,1}=(-52\pi^6-7560\z(3)^2)/945, b_{0,3}=-4\z(5),\,b_{1,2}=-8\z(5)+2\pi^2\z(3)/3,\nonumber\\
&&b_{0,4}=-16\pi^6/1890.\eeqa
where $b_{n,m}$ is symmetric.

Now one can write  $k_1\inn k=k_2.k_3-(k^2)/2$ and $k_2\inn k=k_1.k_3-(k^2)/2$. 
The terms $k^2$ in the vertex \reef{veraatt} will be canceled with the $k^2$ in the denominator of the gauge field propagator producing a bunch of contact terms of one RR and three gauge fields which we are not interested in considering. Neglecting them, one finds the following infinite massless poles 
\beqa
&&-32\pi\mu_p\frac{\eps^{a_{0}\cdots a_{p-1}a}H_{a_{0}\cdots a_{p-1}}}{p!(s+t+u)}\Tr(\lam_1\lam_2\lam_3)
\sum_{n,m=0}^{\infty}\bigg((a_{n,m}+b_{n,m})[s^{m}u^{n}+s^{n}u^{m}]
\bigg\{\bigg[\xi_{3a}(
-2tk_3.\xi_1\nonumber\\&&\times k_3.\xi_2+2uk_3.\xi_1k_1.\xi_2
+2sk_2.\xi_1k_3.\xi_2+us\xi_1.\xi_2)\bigg]+\bigg[3\leftrightarrow 2\bigg]+\bigg[3\leftrightarrow 1\bigg]+\bigg(\bigg[
k_{2a}(-2tk_3.\xi_1\nonumber\\&&\times\xi_2.\xi_3+2uk_1.\xi_2\xi_1.\xi_3+2sk_2.\xi_1\xi_3.\xi_2-2u\xi_1.\xi_2k_1.\xi_3)\bigg]
+\bigg[2\leftrightarrow 1\bigg]\bigg)\bigg\}.
\label{amphigh8}\eeqa 
As a check of our calculations let us compare the above amplitude with the massless poles in  \reef{47} for some values of $n,m$. For $n=m=0$, the amplitude \reef{amphigh8} has the following numerical factor
\beqa
-8(a_{0,0}+b_{0,0})&=&-8(\frac{-\pi^2}{6}+\frac{-\pi^2}{12})=2\pi^2\nonumber\eeqa
A similar term in \reef{47} has the numerical factor  $(4\pi^2c_{0,0})$ which is equal to the above number.  At the order of $\alpha'$, the amplitude \reef{amphigh8} has the following numerical factor
\beqa
-4(a_{1,0}+a_{0,1}+b_{1,0}+b_{0,1})(s+u)&=&0\nonumber\eeqa
 A similar term in \reef{47} is proportional to   $2\pi^2(c_{1,0}+c_{0,1})(s+u)$ which is zero.  At the  order of $(\alpha')^2$, the amplitude \reef{amphigh8} has the following factor 
\beqa
&&-8(a_{1,1}+b_{1,1})su-4(a_{0,2}+a_{2,0}+b_{0,2}+b_{2,0})[s^2+u^2]\nonumber\\
&&=\frac{\pi^4}{3}(2su)+\frac{2\pi^4}{3}(s^2+u^2)
\nonumber\eeqa
A similar term in \reef{47} has the numerical factor  $2\pi^2[c_{1,1}(2su)+(c_{2,0}+c_{0,2})(s^2+u^2)]$ which is equal to the above factor using the coefficients in \reef{577}.  
At the order of $\alpha'^3$, this amplitude has the following factor 
\beqa 
&&-4(a_{3,0}+a_{0,3}+b_{0,3}+b_{3,0})[s^3+u^3]-4(a_{1,2}+a_{2,1}+b_{1,2}+b_{2,1})[su(s+u)]\nonumber\\
&&=-16\pi^2\xi(3)su(s+u)
\nonumber\eeqa
which is equal to the corresponding term in \reef{47}, \ie $2\pi^2[(c_{0,3}+c_{3,0})[s^3+u^3]+(c_{2,1}+c_{1,2})su(s+u)]$.  
At the  order of $(\alpha')^4$, the amplitude \reef{amphigh8} has the following factor 
\beqa
&&-4(a_{4,0}+a_{0,4}+b_{0,4}+b_{4,0})(s^4+u^4)-4(a_{3,1}+a_{1,3}+b_{3,1}+b_{1,3})[su(s^2+u^2)]\nonumber\\
&&-8(a_{2,2}+b_{2,2}) s^2u^2=\frac{4\pi^6}{15}(s^4+u^4+2(s^3u+u^3s)+3s^2u^2)
\nonumber\eeqa
A similar term in \reef{47} has the numerical factor  $2\pi^2[(c_{4,0}+c_{0,4})(s^4+u^4)+(c_{1,3}+c_{3,1})(s^3u+u^3s)+2c_{2,2} s^2u^2]$ which is equal to the above factor using the coefficients in \reef{577}. We have reproduced the known results for terms at ${\cal O}({\alpha'}^4)$$\footnote{Our results up to ${\cal O}({\alpha'}^4)$ are consistent with those terms which found in \cite{Bilal:2001hb} up to on-shell ambiguity and total derivative terms.}$. A similar comparison can be done for all orders of $\alpha'$. Hence, the field theory amplitude \reef{amphigh8} reproduces  exactly the infinite massless poles of the string theory amplitude \reef{47}. This shows that in addition to higher derivative couplings of four gauge fields
 being exact up to zero on-shell, they are also consistent with the momentum expansion of the amplitude $CAAA$.

\subsection{Infinite massless poles for $p=n+2$ case }
The last case is $p=n+2$. Only ${\cal A}_2$ in \reef{48} is non-zero for this case. The trace is calculated as follows 
\beqa
\Tr\bigg(\fsH_{(n)}M_p\Gamma^{dba}
\bigg)&=&\pm\frac{32}{n!}\eps^{a_{0}\cdots a_{p-3}dba}H_{a_{0}\cdots a_{p-3}},
\nonumber\eeqa
Replacing \reef{high} and the above trace in the second part of the amplitude \reef{48}, one finds the electric
part of the amplitude for $p=n+2$ case is given by
\beqa
A^{CAAA}&=&\mp\frac{32}{2(p-2)!}\mu_p \pi^2\Tr(\lam_1\lam_2\lam_3)\eps^{a_{0}\cdots a_{p-3}dea}H_{a_{0}\cdots a_{p-3}}\bigg\{-\sum_{n=-1}^{\infty}{b_n(u+s)^{n+1}\xi_{3d}\xi_{2e}\xi_{1a}}\nonumber\\&&+\bigg(\bigg[\sum_{n=-1}^{\infty}\frac{1}{t}{b_n(u+s)^{n+1}}(-2k_1.\xi_2k_{2e} \xi_{1a}\xi_{3d}-2k_2.\xi_1k_{1a}\xi_{2e}\xi_{3d}-2k_2.\xi_1k_{2a}\xi_{2e}\xi_{3d}
\nonumber\\&&+2k_1.\xi_2k_{1a}\xi_{1e}\xi_{3d}+2\xi_{1}.\xi_{2}k_{1a}k_{2e}\xi_{3d})\bigg]
-\bigg[3\leftrightarrow 1\bigg]-\bigg[3\leftrightarrow 2\bigg]\bigg)
\bigg\}.\label{UI}\eeqa
The amplitude is antisymmetric under the interchange of the gauge fields, so the whole amplitude is zero for the abelian gauge group. The amplitude satisfies the Ward identity for all three gauge fields. Note that only the first term is related to the infinite contact terms while the other terms are indeed infinite massless poles. We are going to analyze all orders of the massless poles in this section and the leading order and next to leading order of the contact terms in the following section.

The amplitudes in s,u and t-channels are very similar, so we 
analyze the amplitude with whole details  only in t-channel. Infinite massless poles in the t-channel should be reproduced in field theory according to the  Feynman diagram of figure 2(a).

\begin{center}
\begin{picture}
(600,105)(0,0)
\Photon(50,105)(100,70){4}{7.5}\Text(75,105)[]{$A_{1}$}
\Photon(50,35)(100,70){4}{7.5}\Text(75,39)[]{$A_{2}$}
\Photon(100,70)(150,70){4}{7.5}\Text(130,88)[]{$A$}
\Gluon(150,70)(200,105){4.20}{5}\Text(170,105)[]{$C_{p-3}$}
\Photon(150,70)(200,35){4}{7.5}\Text(175,39)[]{$A_{3}$}
\SetColor{Black}
\Vertex(100,70){1.5} \Vertex(150,70){1.5}
\Text(130,20)[]{(a)}
\Photon(295,105)(345,70){4}{7.5}\Text(320,105)[]{$A_{1}$}
\Photon(295,70)(345,70){4}{7.5}\Text(315,80)[]{$A_2$}
\Photon(295,35)(345,70){4}{7.5}\Text(315,35)[]{$A_3$}
\Gluon(345,70)(395,105){4.20}{5}\Text(365,105)[]{$C_{p-3}$}
\Vertex(345,70){1.5}
\Text(345,20)[]{(b)}
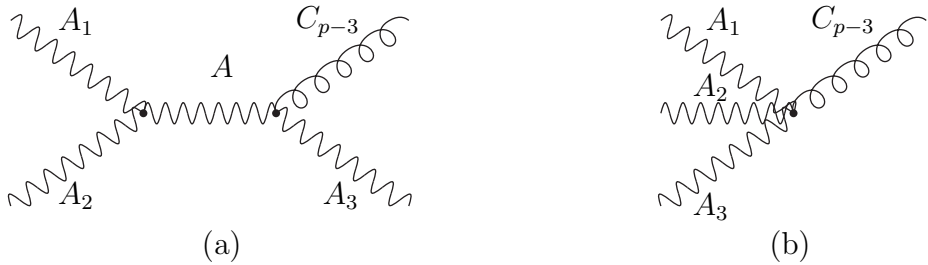
\end{picture}\\ {\sl Figure 2 : The Feynman diagrams (a) and (b) corresponding to the massless pole of the amplitude \reef{UI} and the couplings \reef{UII}.} 
\end{center}
Therefore the infinite massless poles are given by the Feynman amplitude  
\beqa
{\cal A}&=&V^a_{\alpha}(C_{p-3},A_3,A)G^{ab}_{\alpha\beta}(A)V^b_{\beta}(A,A_1,A_2),\labell{amp42}\eeqa
where the vertices and propagator are 
\beqa
V^a_{\alpha}(C_{p-3},A_3,A)&=&\frac{\mu_p(2\pi\alpha')^2}{(p-2)!}\eps^{a_{0}\cdots a_{p-1}a}H_{a_{0}\cdots a_{p-3}}\xi_{3a_{p-2}}k_{a_{p-1}}\Tr(\lam_3\Lambda_\alpha)\sum_{n=-1}^{\infty}b_n(\alpha'k_3.k)^{n+1},\nonumber\\
V^b_{\beta}(A,A_1,A_2)&=&-iT_p(2\pi\alpha')^{2}\Tr(\lam_1\lam_2\Lambda_\beta)\bigg[\xi_1^b(k_1-k).\xi_2+
\xi_2^b(k-k_2).\xi_1+\xi_1.\xi_2(k_2-k_1)^b\bigg],\nonumber\\
G_{\alpha\beta}^{ab}(A)&=&\frac{i\delta_{\alpha\beta}\delta^{ab}}{(2\pi\alpha')^2 T_p
(t)},\nonumber\eeqa
where the propagator is derived from the standard gauge kinetic term arising in the expansion
of the Born-Infeld action. Note that the vertex $V^b_{\beta}(A,A_1,A_2)$ is found from the standard non abelian kinetic term of the gauge field, and also the vertex $V^a_{\alpha}(C_{p-3},A_3,A)$ is found from the higher derivative extension of the WZ coupling $C_{p-3}\wedge F\wedge F$ \cite{Garousi:2007si}. In the above formula $k$ is the momentum of the off-shell gauge field. The important point is that the vertex $V^b_{\beta}(A,A_1,A_2)$ has no higher derivative correction as it arises from the kinetic term of the gauge field. This vertex has already been found in \cite {hatefi:2008}. Substituting them to the amplitude \reef{amp42} becomes
\beqa
{\cal A}&=&\mu_p(2\pi\alpha')^{2}\frac{1}{(p-2)!t}\Tr(\lam_1\lam_2\lam_3)\eps^{a_{0}\cdots a_{p-1}a}H_{a_{0}\cdots a_{p-3}}\xi_{3a_{p-2}} \sum_{n=-1}^{\infty}b_n\bigg(\frac{\alpha'}{2}\bigg)^{n+1}(s+u)^{n+1}\nonumber\\&&\times\bigg[-2k_1.\xi_2k_{1a_{p-1}} \xi_{1a}-2k_1.\xi_2k_{2a_{p-1}}\xi_{1a}+2k_2.\xi_1k_{1a_{p-1}}\xi_{2a}
+2k_2.\xi_1k_{2a_{p-1}}\xi_{2a}\nonumber\\&&-2\xi_{1}.\xi_{2}k_{2a}k_{1a_{p-1}}\bigg].\nonumber\eeqa
which describes exactly the same infinite massless poles of the string theory amplitude \reef{UI} in t-channel.
\subsection{Contact terms}
After finding all infinite massless poles, we now extract the
low energy contact terms of the string amplitude for $p=n+2$ case. Substituting \reef{high} into \reef{UI}, one finds the following contact terms at leading order and next to the leading order
\beqa
A^{CAAA}&=&\mp\frac{32\mu_p \pi^2}{2(p-2)!}\Tr(\lam_1\lam_2\lam_3)\eps^{a_{0}\cdots a_{p-3}dea}H_{a_{0}\cdots a_{p-3}}\bigg\{\xi_{3d}\xi_{2e}\xi_{1a}+\frac{\pi^2}{6}\xi_{3d}\xi_{2e}\xi_{1a}\bigg[(s+u)^2\nonumber\\&&+t(t+2s+2u)\bigg]-\bigg(\bigg[\frac{\pi^2}{6}(t+2s+2u)(-2k_1.\xi_2k_{2e} \xi_{1a}\xi_{3d}-2k_2.\xi_1k_{1a}\xi_{2e}\xi_{3d}\nonumber\\&&-2k_2.\xi_1k_{2a}\xi_{2e}\xi_{3d}
+2k_1.\xi_2k_{1a}\xi_{1e}\xi_{3d}+2\xi_{1}.\xi_{2}k_{1a}k_{2e}\xi_{3d})\bigg]
-\bigg[3\leftrightarrow 1\bigg]\nonumber\\&&-\bigg[3\leftrightarrow 2\bigg]\bigg)
\bigg\}.\label{UII}\eeqa
The first term is reproduced by $CAAA$ coupling of the following  gauge invariant coupling
\beqa
\frac{\mu_p}{2!}(2\pi\alpha')^2\Tr(C_{p-3}\wedge F\wedge F),
\eeqa
which  is given exactly by the WZ terms in \reef{WZ'} after expanding the exponential. The other terms in \reef{UII} should be related to the higher derivative extension of the above coupling. However, there are various higher derivative gauge invariant couplings which make a contribution to the contact terms of the S-matrix element of $CAAA$. Comparing them with the string theory contact terms in \reef{UII}, one can not fix all their coefficients uniquely.  One particular set of higher derivative gauge invariant couplings that  reproduces the contact terms in \reef{UII} is as follows :
\beqa
&&\frac{\mu_p(2\pi\alpha')^2\pi^2}{6(p-3)!}\eps^{a_{0}\cdots a_{p-4}abcd}C_{a_{0}\cdots a_{p-4}}\bigg[-\frac{1}{2}D^{\alpha}D^{\beta}F_{ab}D_{\alpha}D_{\beta}F_{cd}-D_{\alpha}F_{\beta b }D^{\alpha}D_{a}D^{\beta}F_{cd}
\nonumber\\&&+3D_{b}D^{\alpha}D^{\beta}F_{a\alpha }D_{\beta}F_{cd }-\frac{3}{2}D_{a}D^{\alpha}D^{\beta}D_{\beta}F_{b\alpha }F_{cd}
-10D^{\alpha}D^{\beta}D_{d}F_{a\alpha }D_{c}F_{b\beta }\nonumber\\&&-4D_{c}F_{a\alpha }D^{\alpha}D^{\beta}D_{d}F_{b\beta }
-\frac{3}{4}D^{\alpha}D_{\beta}D_{\alpha}D^{\beta}F_{ab}F_{cd}-6D^{\beta}F_{a\alpha }D^{\alpha}D_{c}D_{d}F_{b\beta }\bigg].\labell{699}
\eeqa
where $D_aF=\prt_aF-i[A_a,F]$.
Among the couplings in \reef{699}, only  the first coupling  has non-zero on-shell $CAA$ coupling. This coupling is obtained from the S-matrix element of one RR and two gauge field vertex operators. This coupling has also been used in the previous section to verify that the infinite massless poles  \reef{UI} are reproduced by the higher derivative couplings in field theory. All couplings in \reef{699} are at $(\alpha')^4$ order. The next order terms should be at $(\alpha')^5$ order, and so on.
Hence, the leading order terms of the momentum expansion of the S-matrix element \reef{48} correspond to the Feynman amplitudes resulting from BI and WZ actions and the higher order terms correspond to the higher derivative corrections to the WZ couplings.
Note that the above higher derivative WZ couplings are valid when $p_ap^a\rightarrow 0$. Hence, they can not be compared with `constant RR field'
as a result of  the BSFT.

\section*{Acknowledgment}
The author would like to thank M.R.Garousi, A.Ghodsi, P.Vanhove and I.Y.Park for helpful conversations. He would also like to thank J.Polchinski for comments and A.Fotopoulos for useful suggestions during the winter school
at CERN.  It is a great pleasure to thank  Luis \'Alvarez-Gaum\'e for several valuable discussions. The author also acknowledges  the theory division of CERN for its hospitality. This work was supported by the Ministry of Science, Research and Technology in Iran.


\end{document}